\documentclass[%
 reprint,
 amsmath,amssymb,
 aps,
]{revtex4-2}

\usepackage{graphicx}
\usepackage{xcolor}
\usepackage{dcolumn}
\usepackage{bm}
\usepackage{booktabs}
\newcolumntype{C}[1]{>{\centering\let\newline\\\arraybackslash\hspace{0pt}}m{#1}}

\begin{document}


\title{Indirect Tunneling Enabled Spontaneous Time-Reversal Symmetry Breaking and Josephson Diode Effect in TiN/Al$_2$O$_3$/Hf$_{0.8}$Zr$_{0.2}$O$_2$/Nb tunnel junctions}

\author
{Shaoqing Ding$^{1,2}$, Jinyuan Yao$^{1,2}$, Zhen Bi$^{1,2}$, Quyen Tran$^{2,3}$, Bangzhi Liu$^{2}$, \\Qi Li$^{1,2}$, Susan Trolier-McKinstry$^{2,4}$, Thomas N. Jackson$^{2,3}$, Ying Liu$^{1,2}$}

\thanks{Corresponding author. Email: yxl15@psu.edu}

\affiliation{$^{1}$Department of Physics, The Pennsylvania State University, University Park, Pennsylvania 16802, USA\\
$^{2}$Materials Research Institute, The Pennsylvania State University, University Park, Pennsylvania 16802, USA\\
$^{3}$School of Electrical Engineering and Computer Science, The Pennsylvania State University, University Park, Pennsylvania 16802, USA\\
$^{4}$Materials Science and Engineering Department, The Pennsylvania State University, University Park, Pennsylvania 16802, USA}

\date{\today}

\begin{abstract}
Josephson diode (JD) effect in Josephson tunnel junctions (JTJs) has attracted a great deal of attention due to its importance for developing superconducting-circuitry-based quantum technologies. Even though the preparation of high-quality JTJs by techniques employed in semiconductor industry has been demonstrated, which was an important milestone because JTJs are the building blocks of superconducting electronics even before the quantum era, the JD effect has not been accomplished in them, neither has the highly desirable electrical control of the effect. We report here the fabrication of JTJs featuring a composite tunnel barrier of Al$_2$O$_3$ and Hf$_{\mathrm{0.8}}$Zr$_\mathrm{0.2}$O$_2$ using complementary-metal-oxide-semiconductor (CMOS) compatible atomic layer deposition (ALD). These JTJs were found to show the JD effect in nominally zero magnetic fields with nonreciprocity controllable via an electric training current, yielding a surprisingly large diode efficiency. The quasiparticle tunneling, through which the Josephson coupling in a JTJ is established, was found to show theoretically expected gap features but no nonreciprocity. We attribute these observations to the simultaneous presence of positive and negative local Josephson couplings in the JTJs, with the negative Josephson coupling originating from indirect tunneling, which results in spontaneous time-reversal symmetry breaking. The double-minima washboard potential for the ensemble-averaged phase difference in the RCSJ (resistively and capacitively shunted junction) model is shown to fully account for the experimentally observed JD effect. 
\end{abstract}

\maketitle

\section{Introduction}
Josephson tunnel junctions (JTJs)\cite{josephson_possible_1962,josephson_coupled_1964,anderson_probable_1963,ambegaokar_tunneling_1963} comprising two superconducting electrodes separated by a insulating tunnel barrier have played a central role in developing superconducting electronics crucial for the development of superconducting quantum technologies, including quantum computation. The critical current of a JTJ, $I_\mathrm{c}$, is typically independent of the direction of the current. However, JTJs can also exhibit nonreciprocal $I_\mathrm{c}$ values, potentially providing functionalities similar to those of voltage-biased semiconductor diodes\cite{shockley_theory_1949}. A nonreciprocal JTJ is also known as a Josephson diode (JD), a device concept proposed originally to take advantage of the availability of electron- and hole-doped superconductor that can be placed in close proximity to create a depletion layer similar to that in a semiconductor PN junction\cite{hu_proposed_2007}. However, in recent years any Josephson junction or even a single superconductor exhibiting nonreciprocal behavior is also referred to as JD. Instead, the nonreciprocity in the nonreciprocal JTJ is to arise from the breaking of the space-inversion and time-reversal symmetries (SIS and TRS) without the formation of the depletion layer. The JD effect has been found in a wide range of single Josephson junctions (as opposed to other superconducting devices or materials) consisting of a magnetic barrier \cite{jeon_zero-field_2022, qiu_emergent_2023,trahms_diode_2023}, layered van der Waals materials \cite{wu_field-free_2022,chiles_nonreciprocal_2023,diez-merida_symmetry-broken_2023,qiu_emergent_2023,zhao_time-reversal_2023,ghosh_high-temperature_2024,kim_intrinsic_2024,zhang_magnetic-field-free_2024}, topological materials \cite{pal_josephson_2022,li_interfering_2024}, or a superconducting weak link \cite{trimble_josephson_2021,turini_josephson_2022,matsuo_josephson_2023,gupta_gate-tunable_2023,ciaccia_gate-tunable_2023,baumgartner_supercurrent_2022,costa_sign_2023,gupta_gate-tunable_2023,coraiola_flux-tunable_2024,zhang_magnetic-field-free_2024}, including that utilizing a vortex trap\cite{golod_demonstration_2022}.  

It is often undesirable if the JD effect requires the presence of external magnetic field to break the TRS explicitly from a technological standpoint because of the undesired challenges in controlling superconducting vortices, motivating the search for magnetic-field-free JD effect\cite{wu_field-free_2022,golod_demonstration_2022,jeon_zero-field_2022,chiles_nonreciprocal_2023,zhang_magnetic-field-free_2024,zhao_time-reversal_2023}. A ferroelectric JTJ was proposed to feature the JD effect\cite{trimble_josephson_2021,zhang_general_2022}, which will also serve as a platform for exploring other physical phenomena originating from the interplay between superconductivity and ferroelectricity including the presence of exotic collective modes\cite{ktitorov_josephson_1994}. 

The electrical control of the JD effect, which has so far been rarely achieved\cite{golod_demonstration_2022,zhao_time-reversal_2023}, is also crucial for the applications of JD. Only three studies reported magnetic-field-free JD effect in single JTJs (excluding non-tunneling or more complex devices), among which only one demonstrated the electrical control of the effect (Table\,\ref{tab:1}). In addition, it is also desirable that JDs are fabricated using materials and processes compatible with the complementary metal-oxide semiconductor (CMOS) technology in view of its dominance in the current infrastructure anchoring semiconductor manufacturing. However, the fabrication of JDs by CMOS compatible techniques has not been demonstrated thus far. The use of CMOS-compatible atomic layer deposition (ALD) to prepare JTJs, which was attempted previously but not to achieve the JD effect\cite{lu_fabrication_2013,wilt_atomically_2017,huang_hafnium_2013}, is highly desirable.

\begin{table}
\centering
\caption{\label{tab:1} \textbf{Summary of reported nonmagnetic JTJs showing the JD effect in zero magnetic fields.} BE and TE are bottom and top electrodes, TB: tunneling barrier, C: controllability, $\eta$: diode efficiency defined by $\eta=\big|(I\mathrm{_c^+}-I\mathrm{_c^-})/(I\mathrm{_c^+}+I\mathrm{_c^-})\big|$, and BSCCO: Bi$_2$Sr$_2$CaCu$_2$O$_{8+x}$. \\
}
\begin{tabular}{C{1.5cm}C{1.5cm}C{1.5cm}C{1cm}C{0.9cm}C{1.4cm}}
\toprule
BE & TE & TB & $\eta$ & C & Ref.\\
\hline
NbSe$_2$ & NbSe$_2$ & 2D crystal & 0.08 & no & \cite{wu_field-free_2022}
\\

BSCCO & BSCCO & natural & 0.07 & no &  \cite{zhu_presence_2021}
\\

BSCCO & BSCCO & natural & 0.18 & yes & \cite{zhao_time-reversal_2023}
\\

TiN & Nb & composite & 0.39 & yes & this work
\\
\bottomrule
\end{tabular}

\end{table}

We report in this article the successful preparation of TiN/Al$_2$O$_3$/Hf$_{0.8}$Zr$_{0.2}$O$_2$/Nb JTJs, which, unexpectedly, exhibit the JD effect with a diode efficiency surpassing previously reported values for nonmagnetic JTJs working in nominally zero magnetic fields (Table\,\ref{tab:1}). We employed a CMOS-compatible thermal and plasma enhanced ALD processes for the growth of the bottom electrode of TiN, the tunneling barrier, and the insulating layer defining the junction area. The use of Zr-doped HfO$_2$, a high-$\kappa$ material, is motivated by the demonstration of ferroelectricity in this material down to such a small thickness\cite{lee_unveiling_2021,gao_identification_2021,cheema_emergent_2022} that makes the realization of a ferroelectric JTJ a realistic possibility. 


\begin{figure}
\includegraphics[width=0.48\textwidth]{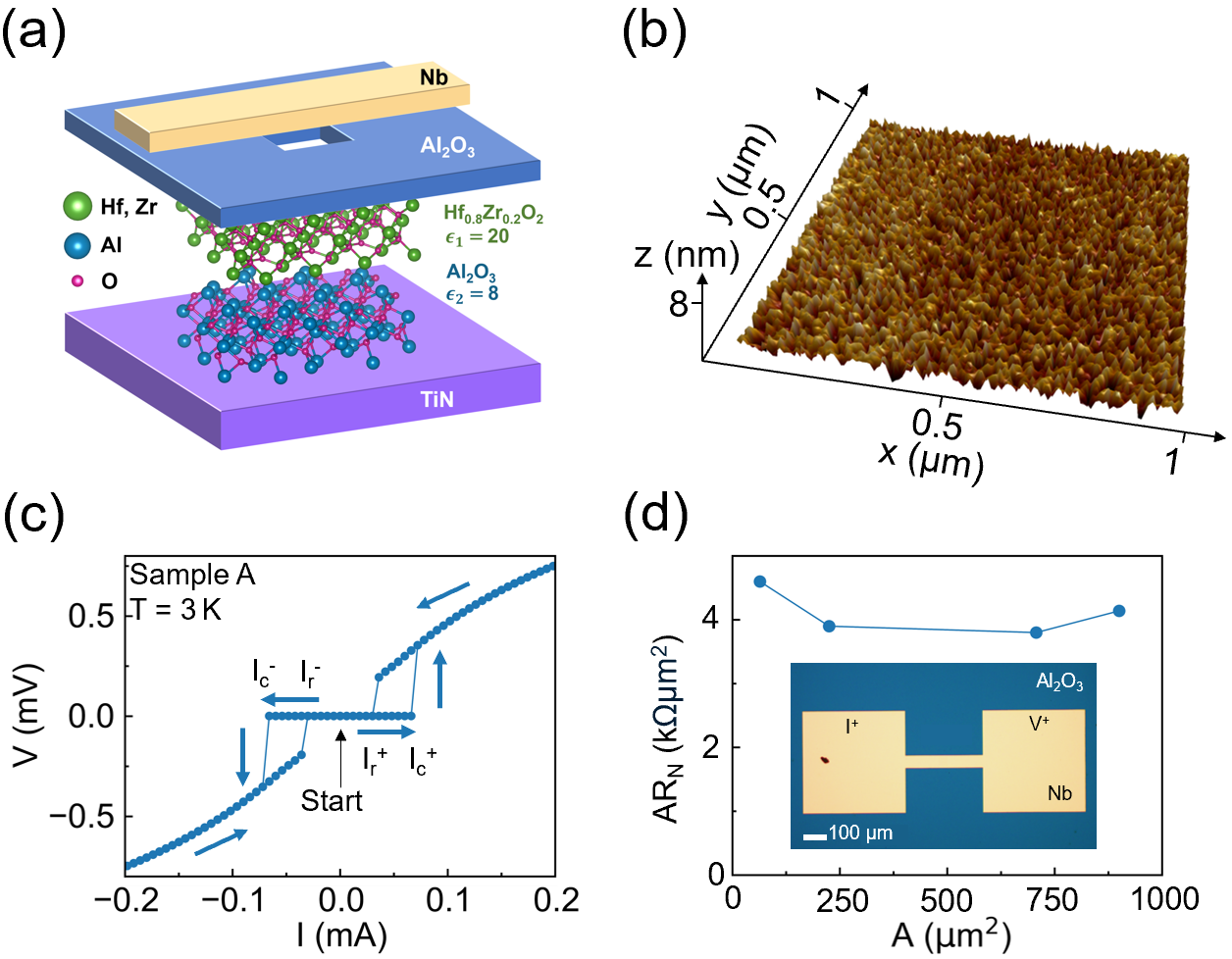}
\caption{\label{fig:1} \textbf{JTJ design and characterization.} 
(a) Schematic of JTJ consisting of 30\,nm TiN, 0.5\,nm Al$_2$O$_3$, 0.7\,nm Hf$_{0.8}$Zr$_{0.2}$O$_2$, and 30\,nm Al$_2$O$_3$, all deposited sequentially in the same ALD chamber. The 100\,nm layer of Nb was sputter-deposited onto the window opened in the top layer of Al$_2$O$_3$ by wet etching.
(b) AFM imaging of a 1 µm$\times$1 µm region showing the smoothness of the bottom layer of TiN.
(c) $I$-$V$ curve at 3\,K and nominally zero magnetic fields with $I_\mathrm{c}^{+,-}$ and $I_\mathrm{r}^{+,-}$ indicated.
(d) Values of $AR_\mathrm{N}$ with an optical image of the device shown in the inset.
}
\end{figure}

\section{CMOS-compatible fabrication of JTJs and characterization}

Size scaling in microelectronics, which for decades has relied on shrinking the physical dimension to increase the density of transistors, appears to have or nearly have met the physical limits in miniaturization. Countless innovations including the 3D integration had to be made to enable the continuation of the scaling laws\cite{salahuddin_era_2018}. The ability of ALD in conformal coating of non-planar semiconductors with atomically thin films of a wide range of materials has made it an indispensable tool in CMOS technology. It is interesting to note that ALD has nevertheless rarely been used in the preparation of JTJs -- only a couple of attempts have been reported thus far \cite{lu_fabrication_2013,wilt_atomically_2017,huang_hafnium_2013}. 

Taking advantage of thermal ALD and plasma-enhanced ALD (PEALD) techniques, we pursued the preparation of JTJ involving a TiN bottom electrode and a composite tunneling barrier of Al$_2$O$_3$/Hf$_{0.8}$Zr$_{0.2}$O$_2$. The barrier was covered by a thick layer of Al$_2$O$_3$, also grown by ALD, with windows etched into the top Al$_2$O$_3$ layer to define junctions of varying sizes. The top electrode of Nb was prepared by sputtering that tends to yield a film and an interface with Hf$_{0.8}$Zr$_{0.2}$O$_2$ of superior quality. The Al$_2$O$_3$ layer was employed to prevent the oxidation of TiN, as interdiffusion of TiN and Hf$_{0.8}$Zr$_{0.2}$O$_2$ is expected due to the high growth temperature (250\,°C) required for the latter\cite{baik_electrical_2017,filatova_control_2018}, and to engineer a dielectric interface featuring a large mismatch in dielectric constants ($\kappa_\mathrm{Al_2O_3}$=8, $\kappa_\mathrm{HZO}$=20). When an electric field is applied across the interface, bound charge accumulates at the interface, which may remain at the interface even after the field is removed. Importantly, similar to the surface bound charge found in a ferroelectric layer, the bound interface charge will also attract (or even capture) mobile charge of opposite sign. The design and the optical image are shown in Fig.\,\ref{fig:1}(a) and the inset of Fig.\,\ref{fig:1}(d) respectively. Electrical contacts to the junction were made by pressing small indium dots onto the top and bottom electrodes (not shown), enabling standard four-point electrical transport measurements at low temperatures.

The atomic force microscopy (AFM) was used to assess the smoothness of the bottom electrode, a 30-nm-thick TiN film, revealing excellent surface smoothness (Fig.\,\ref{fig:1}(b)). The presence of the Josephson coupling in the junction is evident through the observation of the zero-voltage supercurrent in the current ($I$)-voltage ($V$) curves  (Fig.\,\ref{fig:1}(c)). As the current is ramped up from zero, a sharp rise in voltage marks the critical current, $I_{\mathrm{c}}^{+}$. After reaching a preset maximum value, the current is ramped back down to zero, with the voltage returning to zero at the retrapping current, $I_{\mathrm{r}}^{+}$. A similar behavior occurs during the negative current sweep, yielding the corresponding critical and retrapping currents, $I_{\mathrm{c}}^{-}$ and $I_{\mathrm{r}}^{-}$. The large difference between the critical and retrapping currents suggests that the Josephson junction belongs to the underdamped category, ruling out the presence of metallic shorts in the tunneling barrier\cite{tinkham_introduction_2004}. 


The normal-state junction resistance, $R_\mathrm{N}$, can be estimated from the voltage right above the critical current in the $I$-$V$ curve. The contribution to the junction resistance from the TiN film (connected in series with the junction) was found to be significant when the bottom electrode (TiN) becomes non-superconducting because it covers the entire substrate. Using the $I$-$V$ curve to compute $R_\mathrm{N}$ avoids this problem, as the critical current of the junction will be much smaller than that of the large TiN film. To assess the uniformity of the tunnel barrier, values of $AR_\mathrm{N}$, where $A$ is the junction area and $R_\mathrm{N}$ is obtained at 2\,K, for four junction sizes were shown in Fig.\,\ref{fig:1}(d), revealing only small variation for $A$ = 64, 225, 707, and 900\,µ$\mathrm{m^2}$, which indicates again that the tunnel barrier is reasonably uniform and free of metallic shorts.
Values of the superconducting energy gap for the top and bottom electrodes can be estimated from the T$_\mathrm{c}$ values using the BCS theory ($\Delta_{T=0}\approx$1.76$k_\mathrm{B}T_\mathrm{c}$) \cite{tinkham_introduction_2004}, yielding two zero-temperature gap values, $\Delta_\mathrm{Nb}$ = 1.25\,meV for Nb and $\Delta_\mathrm{TiN}$ = 0.52\,meV for TiN, consistent with those found in the literature \cite{townsend_investigation_1962,escoffier_anomalous_2004}.

\section{Quasiparticle tunneling}

\begin{figure*}
\includegraphics[width=\textwidth]{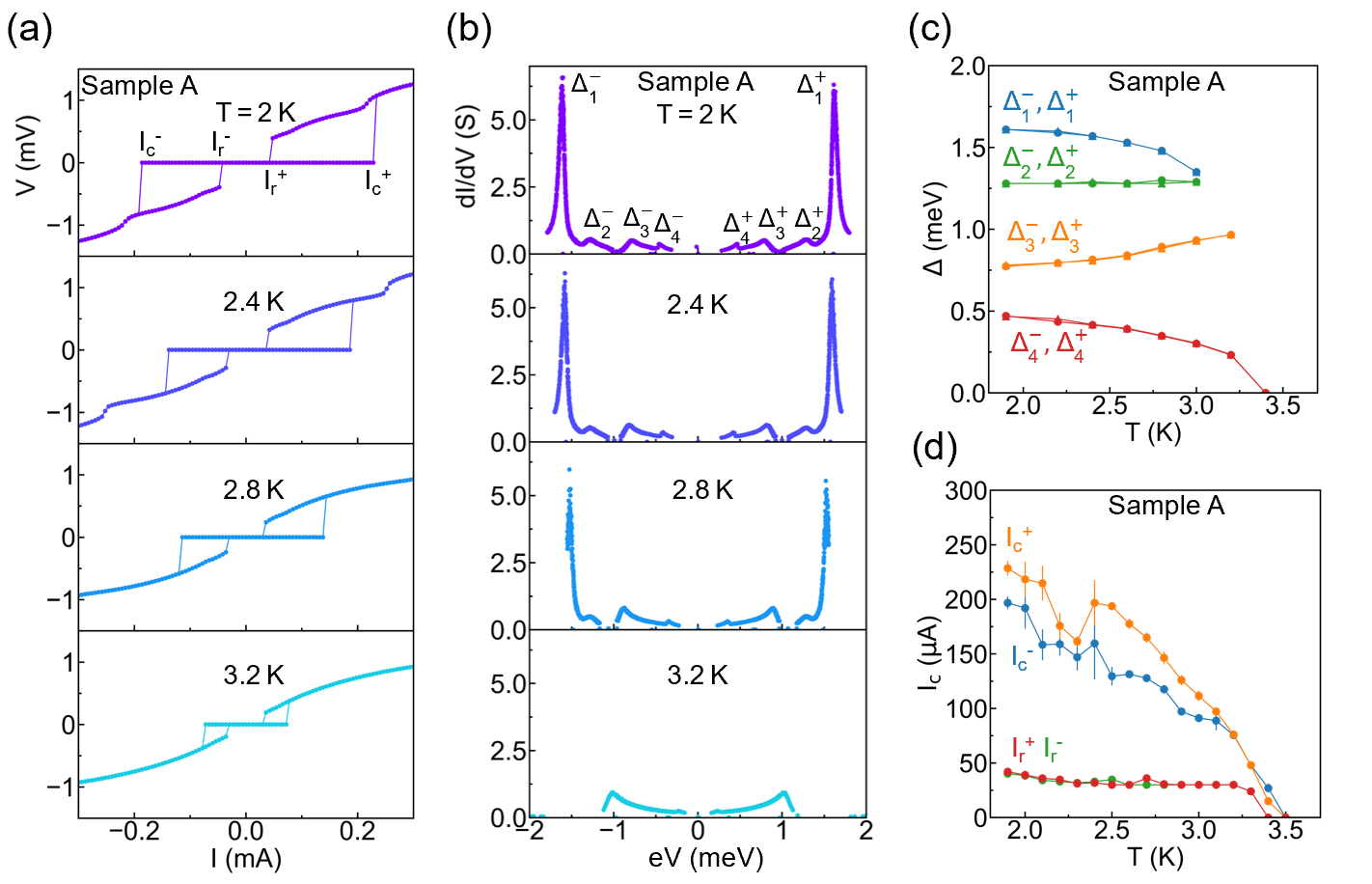}
\caption{\label{fig:2} \textbf{Quasiparticle tunneling and Josephson effect.}
(a) $I$-$V$ curves measured at various temperatures with critical and retrapping currents, $I_\mathrm{c}^{+,-}$ and $I_\mathrm{r}^{+,-}$, indicated. 
(b) d$I$/d$V$ $vs$. $eV$ at various temperatures showing four peaks denoted by $\Delta_\mathrm{1}$, $\Delta_\mathrm{2}$, $\Delta_\mathrm{3}$, and $\Delta_\mathrm{4}$. These features are related to superconducting energy gaps of superconducting electrodes. 
(c) Values of the gap features with those obtained from the negative voltages marked by circles and those obtained from the positive voltages by triangles showing no nonreciprocal behavior.
(d) $I_\mathrm{c}^{+,-}$ and $I_\mathrm{r}^{+,-}$ as a function of temperature without the application of training current. The error bars were obtained from the standard deviation (see text) with nonreciprocal behavior seen clearly.
}
\end{figure*}

The quasiparticle tunneling spectra, d$I$/d$V$ $vs.$ $eV$ curves, provide insights into the nature of the Josephson coupling as the Josephson currents in a JTJ are derived from the phase coherent quasiparticle tunneling (at zero bias). Fig.\,\ref{fig:2}(a) and (b) show the $I$-$V$ and d$I$/d$V$ $vs$. e$V$ curves at selected temperatures are shown. Four peaks marked by $\Delta_\mathrm{1}$, $\Delta_\mathrm{2}$, $\Delta_\mathrm{3}$, and $\Delta_\mathrm{4}$ are clearly seen in d$I$/d$V$ $vs$. e$V$ curves. Theoretically, quasiparticle tunneling between TiN and Nb should yield two peaks in the spectrum, at $\Delta_\mathrm{Nb}$ + $\Delta_\mathrm{TiN}$ and $\Delta_\mathrm{Nb}$ - $\Delta_\mathrm{TiN}$, respectively, through direct tunneling. As stated above, the $T=0$ BCS gap values of Nb and TiN are $\Delta_\mathrm{Nb}=1.38$\,meV and $\Delta_\mathrm{TiN}=0.54$\,meV 
so that $\Delta_\mathrm{1}$ = $\Delta_\mathrm{Nb}$ + $\Delta_\mathrm{TiN}$ and $\Delta_\mathrm{3}$ = $\Delta_\mathrm{Nb}$ - $\Delta_\mathrm{TiN}$. Indeed, the temperature dependences of these two gap features (Fig.\,\ref{fig:2}(b) are as expected\cite{tinkham_introduction_2004}. 

The observation of two other peaks, $\Delta_\mathrm{2}$ and $\Delta_\mathrm{4}$, seen at voltages corresponding to the energy gaps of the two superconducting electrodes, $\Delta_\mathrm{Nb}$ and $\Delta_\mathrm{TiN}$, is not expected theoretically from direct quasiparticle tunneling between two superconductors\cite{tinkham_introduction_2004}. However, such features were observed experimentally\cite{taylor_excess_1963} and explained theoretically in a picture involving direct tunnelings of two quasiparticles at a finite bias voltage\cite{schrieffer_two-particle_1963}. The gap values shown in Fig.\,\ref{fig:2}(c) were found to depend on the temperature as expected from the BCS theory\cite{tinkham_introduction_2004}. The superconducting energy gap decreases as temperature increases but only appreciably when the temperature approaches to roughly 70\% of $T_\mathrm{c}$ is approached (The gap vanishes at $T_\mathrm{c}$). Since the $T_\mathrm{c}$ of Nb is measured to be 8.3\,K, the change of $\Delta_\mathrm{2}$ is expected to be small below 3.5\,K. However, the $T_\mathrm{c}$ of TiN and of the junction is at 3.4\,K, therefore $\Delta_\mathrm{4}$ is expected to change significantly around 2.5\,K. Interestingly, because the gap value starts to drop appreciably when the temperature is raised to about 0.7$T_\mathrm{c}$ and the $T_\mathrm{c}$ values of Nb and TiN are so different, the value of $\Delta_\mathrm{3}$ = $\Delta_\mathrm{Nb}$ - $\Delta_\mathrm{TiN}$ is seen to show an upturn from 2.1 to 3.2\,K, as expected. The observations of the four gap features and the respective temperature dependences of them demonstrate that the tunnel barrier are pinhole-free even though the thickness varies spatially (which makes the double quasiparticle tunneling through thinnest and single one through less thin parts of the tunnel barrier competitive). This in turn demonstrates the power of ALD in the preparation of atomically thin tunnel barrier.

The simultaneous measurements of the $I_\mathrm{c}$ and $R_\mathrm{N}$ values on the TiN/Al$_2$O$_3$/Hf$_{0.8}$Zr$_{0.2}$O$_2$/Nb junctions provide a means to infer the nature of the Josephson coupling. According to the Ambegaokar-Baratoff (A-B) theory\cite{ambegaokar_tunneling_1963}, the $I_\mathrm{c}R_\mathrm{N}$ value is determined by the superconducting energy gaps of the two electrodes and the normal-state junction resistance. The zero-temperature $I_\mathrm{c}$, calculated analytically from the A-B formula using the previously determined gap values, is roughly 1.1\,mV/$R_\mathrm{N}$, which is close to the experimental values ($I_\mathrm{c}R_\mathrm{N}\sim$1.0\,mV) obtained at the lowest measured temperatures. 

\section{Training of the JD effect}

Multiple $I$-$V$ curve loops taken consecutively revealed variations in the critical current while that of the retrapping current was found to vary very little. To quantify the variation in the critical current, five to ten $I$-$V$ loops were measured consecutively. The $I$-$V$ curve loops shown in Fig.\,\ref{fig:2}(a), obtained at various fixed temperatures are those with the $I_\mathrm{c}^{+,-}$ values closest to the average ones. 

The averaged values of $I_\mathrm{c}^{+,-}$ and $I_\mathrm{r}^{+,-}$ shown in Fig.\,\ref{fig:2}(d) reveal clearly nonreciprocal behavior without the application of training current. It is seen that the nonreciprocal behavior decreases as the temperature increases and disappears above around 3.2\,K. A dip in $I_\mathrm{c}^{+}$ is also seen around 2.3\,K. The temperature independence seen in $I_\mathrm{r}^{+,-}$, which is very different from that seen in $I_\mathrm{c}^{+,-}$, can be easily explained if the damping is dominated by high-frequency contributions rather than the frequency independent normal state junction resistance\cite{johnson_effect_1990}, which is likely to be the case in our experimental setup.


\begin{figure*}
\includegraphics[width=\textwidth]{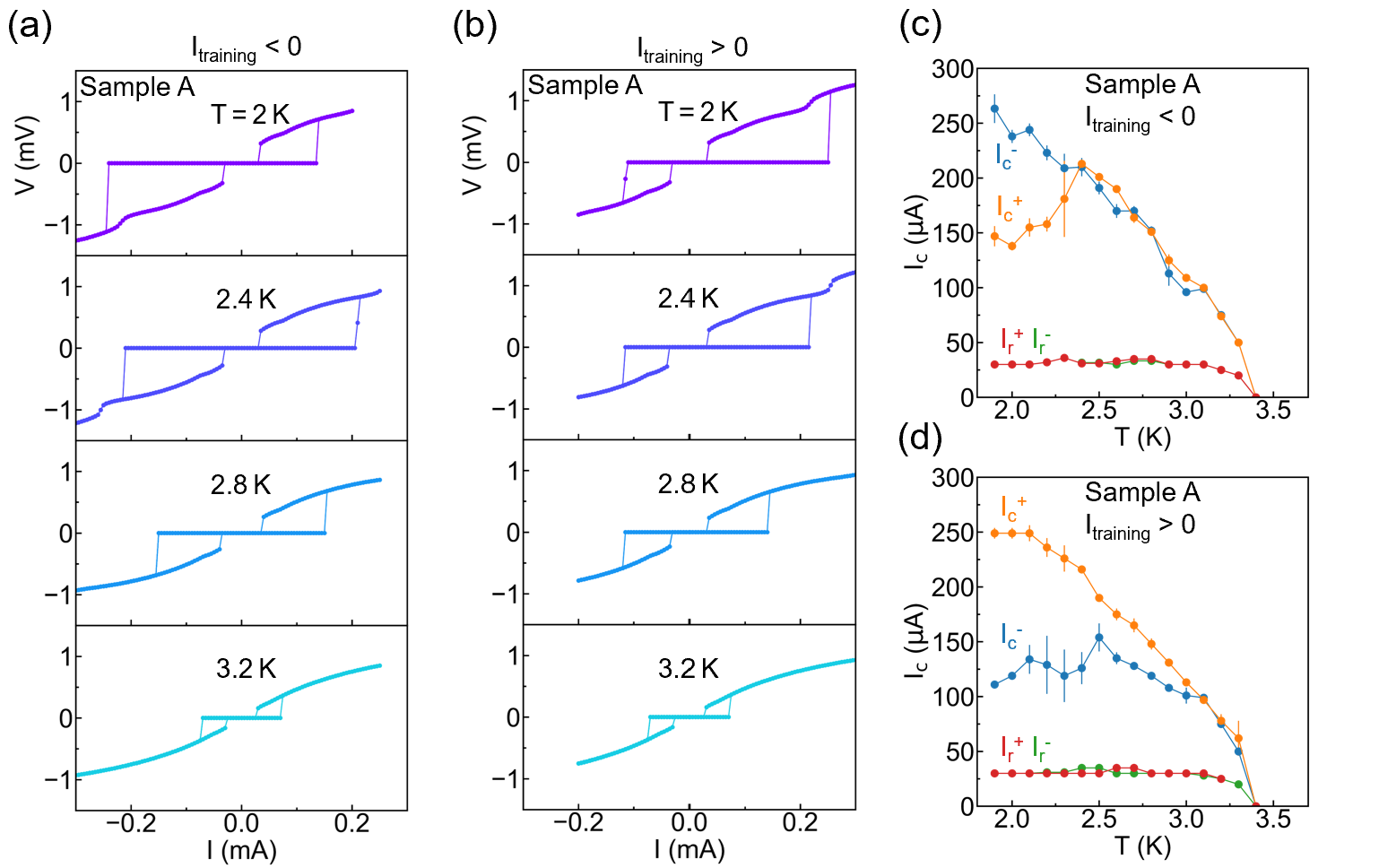}
\caption{\label{fig:3} \textbf{Effect of training by a large electric current.}
(a, b) $I$-$V$ curves at various temperatures as indicated after training using a negative (a) and positive (b) electric current.
(c, d) $I_\mathrm{c}^{+,-}$ and $I_\mathrm{r}^{+,-}$ as a function of temperature after training using a negative (c) and positive (d) electric current. The error bars represent the standard deviation. The polarity of the nonreciprocal behavior is seen to depend on the polarity of the training current.}
\end{figure*}

The JD effect was found to be controllable by a normal-state electric current approximately five times the critical current, which corresponds to an electric field on the order of $10^5$\,V/cm. This ``training" was performed in one of the two ways. The current can be applied at a low temperature, held for five minutes, and then turned off to allow the sample to return to the superconducting state. Alternatively, it could also be applied at 10\,K (above the T$_\mathrm{c}$ of Nb and TiN), held steady while the sample was cooled to the measurement temperature, and then switched off so that the junction returned to the superconducting state. In either way the junction consistently was found to exhibit a larger critical current in the same direction of that of the training current, independent of the original polarity of the JD effect. It is likely that the use of a training current smaller than five times the critical current, which was not attempted in the current experiment, can achieve the same effect. 

In Fig.\,\ref{fig:3}(a) and (c), it is shown that $I_\mathrm{c}^-$ is almost twice as large as $I_\mathrm{c}^+$ after applying a negative training current of -1.2\,mA. After training with a positive current (+1.2\,mA), the value of $I_\mathrm{c}^-$ was greatly reduced and $I_\mathrm{c}^+$ became almost twice as large as $I_\mathrm{c}^-$ (Fig.\,\ref{fig:3}(b) and (d)). After the training, the diode efficiency was significantly enhanced. On the other hand, the effect of training disappears if the sample is warmed up above 10\,K. Similar behavior in training was seen in additional samples featuring reduced diode efficiencies. Interestingly, the critical current flowing in the same direction as the training current appears to be unaffected or even slightly enhanced with its value close to the A-B limit. On the other hand, the critical current in the direction opposite to the training current is significantly suppressed. The different diode effect onset temperatures for sample\,A and B, as shown in Fig.\,3(c, d), may be due to differences in the spatial and energetic distribution of the localized states in the tunnel barrier, as will be discussed later. The retrapping current shows no diode effect, as shown in Fig.\,3(c, d), most likely because the high-frequency damping originating from the leads dominates. Furthermore, the training was found to fix the polarity of the JD as long as the JTJ is kept at low temperatures and below the larger critical current, which makes the device promising for nonvolatile memory in superconducting circuits. 

\section{Indirect tunneling and negative Josephson coupling}

In addition to promising aspects for quantum applications, these TiN/Al$_2$O$_3$/Hf$_{0.8}$Zr$_{0.2}$O$_2$/Nb tunnel junctions studied in the current work have also raised fundamental physical questions. To begin with, the critical current was found to be reduced by the training in one direction while both the $R_\mathrm{N}$ and the gap values are not affected. Moreover, the $I_\mathrm{c}R_\mathrm{N}$ value in the unsuppressed direction remains close to the A-B value. All this suggests that the origin of the observed JD effect is associated with a mechanism that reduces the critical current without changing the normal-state junction resistance and gap values of the two electrodes. It was recognized long ago that the quasiparticle tunneling involving spin flips, not considered in the original A-B theory, would lead to the reduction of the $I_\mathrm{c}R_\mathrm{N}$ value from the A-B value\cite{kulik_magnitude_1966}, which furthermore leads to negative Josephson coupling. Subsequent studies identified several possible mechanisms for negative Josephson coupling, such as magnetic impurities in the tunnel barrier\cite{ shiba_superconducting_1969,bulaevskii_superconducting_1977,glazman_resonant_1989}. However, all these mechanisms require a Hamiltonian that explicitly breaks TRS, making it irrelevant to the current study. 

\begin{figure}
\includegraphics[width=0.45\textwidth]{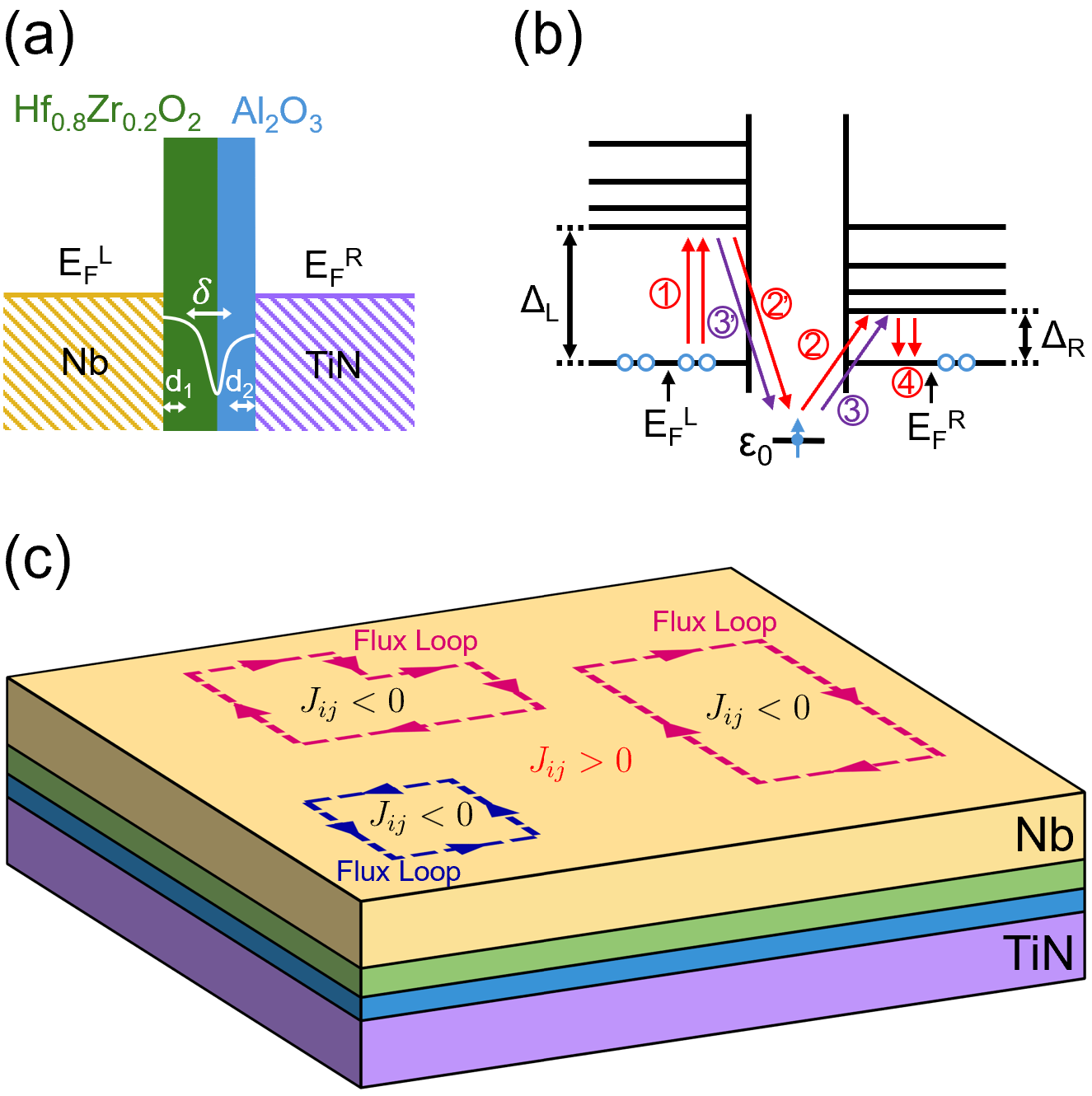}
\caption{\label{fig:4}\textbf{Indirect Josephson tunneling and spontaneous TRS breaking.} 
(a) Schematic of a JTJ featuring a singly occupied localized state in the tunnel barrier. $E_F^{L,R}$ is the Fermi energies of the left and right electrode, $\delta$ is the spatial extension of the localized state, and the total barrier thickness is $d$ = $d_1$ + $d_2$ + $\delta$. 
(b) Indirect Josephson tunneling through the localized state with an energy of $-\varepsilon_0<0$ below the aligned Fermi energies. Cooper pairs in the condensates are shown by open circles but quasiparticles in the excited states are not shown. Intermediate processes to transfer the Cooper pair one quasiparticle at a time through the barrier are indicated by circled numbers, including the simultaneous tunnelings of the electron at the localized state to the right and a quasiparticle on the left to the localized state (2, 2' and 3, 3'). The two quasiparticles arriving at the right superconducting electrode must be interchanged in ordering before they condense into the condensate (4), resulting in a negative Josephson coupling.
(c) Clockwise and counterclockwise magnetic flux loops (thick dash lines) formed along the boundary of clusters of mini Josephson junctions featuring a negative Josephson coupling, breaking the TRS spontaneously.
}
\end{figure}

On the other hand, a negative Josephson coupling may emerge in the JTJs studied in the current work through previously identified indirect tunneling processes \cite{spivak_negative_1991,kivelson_aharonov-bohm_1992} despite TRS not being explicitly broken. Consider now a singly occupied localized state ($-\varepsilon_0 < 0$) within the tunnel barrier (see below for discussion on its origin) with its energy below the aligned Fermi energies of the two superconducting electrodes (Fig.\,\ref{fig:4}(a)). The energy of the localized state as a function of the number of electrons ($n_0 = 0, 1, 2$) in the localized state is $E_{\mathrm{loc}}=-\varepsilon_0 n_0 +U n_0 (n_0-1)/2$. Assuming further that adding a second electron leads to a large Coulomb interaction $U$, making it energetically unfavorable. The full Hamiltonian of the system, including localized states in the barrier, is given by \(H=H_L+H_R+H_T\), where $H_L$ and $H_R$ are BCS reduced Hamiltonians for the two superconductors and $H_T$ represents the tunneling terms. For the direct tunneling, the tunneling Hamiltonian is given by
\begin{equation}
H_{T} = \sum_{kqs}(t_{kq}c^+_{Lks}c_{Rqs}+h.c.)
\end{equation}
 where $c^{+}_{Lks}$, $c_{Lks}$ and $c^{+}_{Rqs}$, $c_{Rqs}$ are the respective creation and annihilation operators and $t_{kq}$ is the tunneling matrix element for removing an electron with momentum $q$ and spin $s$ in the right to become one in the left superconductor with a momentum $k$ and spin $s$, respectively. The tunneling matrix can be estimated in the WKB approximation,
\begin{equation}
t_{ks}\sim\exp{(-\frac{1}{2\hbar}\sqrt{2m_eU}d)}
\end{equation}
where $U$ is a parameter characterizing the height and $d$ the thickness of the tunneling barrier. Both Josephson and quasiparticle tunneling currents have been shown in the previous section where the four peaks in the tunneling spectra are expected.

The Hamiltonian for the indirect tunneling is given by
\begin{equation}
H_T=\sum_{ks}(T_{Lk}c^+_{Lks}c_{0s}+h.c.)+\sum_{qs}(T_{Rq}c^+_{Rqs}c_{0s}+h.c.)
\end{equation}
where $T_{Lk}$ and $T_{Rq}$ are tunneling matrix elements and $c_{0s}$ is the annihilation operator for the localized state.
In the WKB approximation, the tunneling matrix elements are given by 
\begin{subequations}
\begin{equation}
T_{Lk}\sim\exp{(-\frac{1}{2\hbar}\sqrt{2m_eU_1}d_1)}
\end{equation}
\begin{equation}
T_{Rq}\sim\exp{(-\frac{1}{2\hbar}\sqrt{2m_eU_2}d_2)}
\end{equation}
\end{subequations}
where $U_1$ and $U_2$ are parameters characterizing the heights of the tunneling barriers, $d_1$ and $d_2$ are the thicknesses to the left and the right side of the localized states, respectively, and $\delta$ is the spatial extension of the localized state, which cannot be significantly smaller than the barrier thickness ($d$ = 1.2\,nm) as shown in Fig.\,\ref{fig:4}(a). 

The total thickness of the tunnel barrier is reduced because of the finite size of the localized state, making it plausible that the probability of indirect tunneling is comparable to that of direct tunneling.  
Indirect tunneling at zero bias (Fig.\,\ref{fig:4}(b)) can lead to a negative local Josephson coupling as shown previously\cite{spivak_negative_1991,kivelson_aharonov-bohm_1992} even though neither spin flips\cite{kulik_magnitude_1966} nor magnetic impurities required in the previous studies\cite{ shiba_superconducting_1969,bulaevskii_superconducting_1977,glazman_resonant_1989} are involved. In an experimental sample, more than one localized state will be present and their effects will be added. Naively, the total critical current will drop if the local Josephson couplings in some area of the junction become negative.  

\begin{figure}[htbp!]
\includegraphics[width=0.48\textwidth]{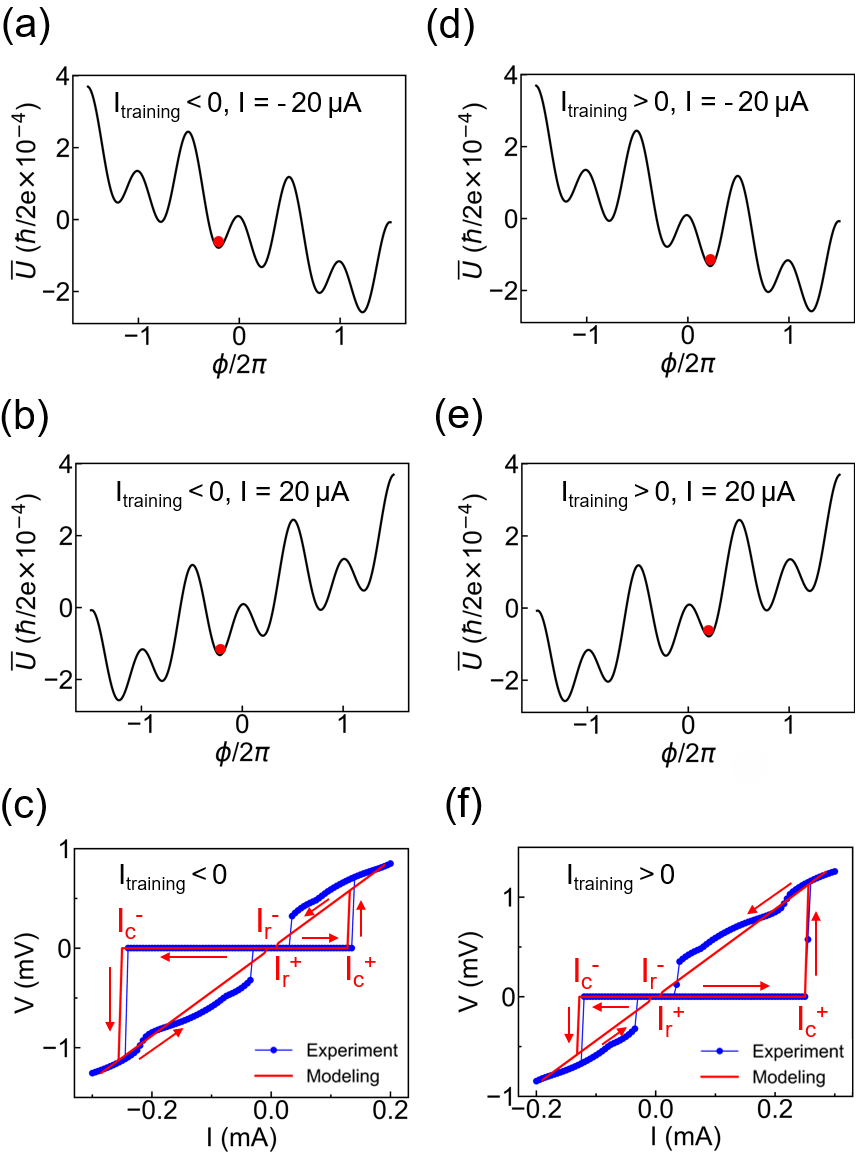}
\caption{\label{fig:5} \textbf{Double minima potential and the JD effect.}
The ensemble averaged potential ($\overline{U}$) in the RCSJ model for the JTJ featuring both the first and the second order terms that results in an energy landscape featuring two distinct minima, as shown in (a, b, d, e). The values of R$_N$, C, J, and J$_2$ used to plot $\overline{U}$ are those of Sample\,A (see text). The total current flowing through the JTJ, $I$, with the direction of $I_\mathrm{training}$ indicated. The effect of training with an $I_\mathrm{training}$ $<$0 ($>$ 0) is assumed to set the initial conditions for Eq.\,7 such that the phase particle (filled circle) rests in a valley to the left (or right) of the higher peak in $\overline{U}$ as indicated.
The $I$-$V$ curves obtained from solving Eq.\,7 for Sample\,A (solid lines) are shown in (c) and (f), along with experimental data. The experimental and calculated values for $I_\mathrm{c}^-$ and $I_\mathrm{c}^+$ are consistent with one another but not those for $I_\mathrm{r}^-$ and $I_\mathrm{r}^+$ because the latter are dominated by the high frequency damping not included in the modeling.  
}
\end{figure}

\section{Spontaneous TRS breaking and electrical control of the JD effect}

The simultaneous presence of positive and negative Josephson couplings in the JTJ will lead to the formation of supercurrent and corresponding magnetic flux loops (Fig.\,\ref{fig:4}(c), which in turn leads to the TRS breaking, a necessary but not sufficient condition for the JD effect.
We note that our junctions are not $\phi_0$-junctions since a phase shift $\phi_0$ in the current phase relation would correspond to explicit rather than spontaneous TRS breaking \cite{buzdin_direct_2008,konschelle_theory_2015,strambini_josephson_2020,jeon_zero-field_2022,kim_intrinsic_2024}. To obtain the nonreciprocal $I_\mathrm{c}$ values, the distribution of the local Josephson coupling and its effect on the critical current of the junction need to be analyzed. Denoting the Josephson junction plane as the $x$-$y$ plane, we write the effective Hamiltonian in the form
\begin{equation}
H_\mathrm{eff}=H_\mathrm{eff}^{\Delta}(|\Delta_L|,|\Delta_R|)+H^{\phi}_\mathrm{eff}[J(x,y),\phi_L(x,y),\phi_R(x,y)]
\end{equation}
which possess the TRS. Note that the amplitudes of the superconducting energy gap remain constant in both superconducting electrodes but their phases vary spatially with the phase rigidity length, which is roughly the zero-temperature superconducting coherence length ($\xi$ $\sim$ 100\,nm). The large JTJ used in the present study can be modeled by a mini JTJ array, with the energy associated with the phases depending on not only the Josephson energy, \(E_J=-\sum_{ij}J_{ij}\cos{(\phi_{L,ij}-\phi_{R,ij}})\) but also the kinetic energy associated with phase variations within the two superconducting electrodes, \(E_{L,R}^\mathrm{\phi}=\sum_{ij}\kappa^{L,R}\xi^{-2}[(\phi^{L,R}_{ij}-\phi^{L,R}_{i-1j})^2+(\phi^{L,R}_{ij}-\phi^{L,R}_{ij-1})^2]\), where the indices ($i$,$j$) mark the position of the mini junction. 

The critical currents of these mini junctions can be obtained in the resistively and capacitively shunted junction (RCSJ) model described by
\begin{subequations}
\begin{equation}
    \Big(\frac{\hbar}{2e}\Big)^2C_{ij}\frac{d^2\phi_{ij}}{dt^2}+\Big(\frac{\hbar}{2e}\Big)^2\frac{1}{R_{\mathrm{N},ij}}\frac{d\phi_{ij}}{dt}=-\frac{dU_{ij}}{d\phi_{ij}}
\end{equation} 
\begin{equation}
    U_{ij}=-J_{ij}\cos{(\phi_{ij})}-\frac{\hbar}{2e}I_{ij}\phi_{ij}
\end{equation} 
\end{subequations}
in which the phase difference across the junction ($\phi_{ij} = \phi_{L,ij}-\phi_{R,ij}$) is treated as a particle moving in the potential energy of $U_{ij}$. $C_{ij}$ and $R_{\mathrm{N},ij}$ are the capacitance and normal-state resistance of the mini junction, respectively. $J_{ij}=(\hbar/2e)I_{c0,ij}$ is the Josephson coupling energy where $I_{c0,ij}$ is the critical current of the mini junction and $I_{ij}$ denotes the total current flowing through the mini junction. 

The critical current of the whole junction cannot be obtained by adding up the critical currents of all mini junctions as phases of these junctions are correlated in a complicated shorting process -- mini junctions with larger critical current will short mini junctions with the smaller critical currents. However, it is reasonable to pursue ensemble averaging of the randomly distributed local Josephson coupling. Following insights provided by previous studies\cite{Zyuzin_theory_2000,zhao_time-reversal_2023, yuan_inhomogeneity-induced_2023, volkov_josephson_2024}, it appears that the physics of such a system can be captured by a second-order term in the potential energy for the ensemble averaged phase difference with the RCSJ model described by 
\begin{subequations}
\begin{equation}
\Big(\frac{\hbar}{2e}\Big)^2C\frac{d^2\phi}{dt^2}+\Big(\frac{\hbar}{2e}\Big)^2\frac{1}{R_\mathrm{N}}\frac{d\phi}{dt}=-\frac{d\overline{U}}{d\phi}
\end{equation} 
\begin{equation}
\overline{U}=-\overline{J}\cos{(\phi)} +J_2\cos{(2\phi)}-\frac{\hbar}{2e}I\phi
\end{equation}
\end{subequations}
where $C$ and $R_\mathrm{N}$ are the capacitance and normal state resistance of the whole junction, respectively. The value of the capacitance of the junction, $C$, can be estimated from the area and the dielectric constants and thicknesses of Al$_2$O$_3$ and HZO and $R_\mathrm{N}$ is an experimentally measured parameter. $\phi=\overline{\phi}_L-\overline{\phi}_R$ is the difference of the ensemble averaged phase. In the potential energy $\overline{U}$, $\overline{J}>0$ is the average Josephson coupling energy (assuming that the positive Josephson coupling dominates), $J_2 > 0$ is the coefficient of the second-order term, and $I$ is the total current flowing through the junction. This second-order term leads to double minima in the tilted washboard potential for the ensemble-averaged phase difference.


Assuming that the training current sets the initial conditions for Eq.\,7 as shown schematically in Fig.\,\ref{fig:5} (a, b, d, e), the equation can be solved numerically to obtain the $I$-$V$ curves. For example, to model the $I$-$V$ curve obtained on Sample\,A at 2\,K (Fig.\,\ref{fig:3} (a, b)), which features junction parameters of $R_\mathrm{N}$ = 4.4\,$\Omega$, $C$ = 82\,pF, and $\overline{J}$ = 176.6\,meV ($\sim$ 205 $\sqrt{\Delta_\mathrm{Nb}\Delta_\mathrm{TiN}}$), taking $J_2$ = 195.6\,meV ($\sim$ 227 $\sqrt{\Delta_\mathrm{Nb}\Delta_\mathrm{TiN}}$) will yield $I$-$V$ curves reasonably close to those obtained experimentally (Fig.\,\ref{fig:5} (c, f)). The substantially large $\overline{J}$ value indicates that positive local Josephson coupling dominates, as expected. 

Exactly how the training current sets up the initial conditions in the RCSJ model? It is unlikely that the training current, which generates an electric field on the order of $10^5$\,V/cm, can displace the localized states in the barrier if the states are generated by, for example, oxygen deficiencies. However, applying such a field will generate bound charge at the Al$_2$O$_3$/Hf$_{0.8}$Zr$_{0.2}$O$_2$ interface with its sign and density determined by the applied field \cite{griffiths_introduction_2023}. It is possible that even after the training current is switched off, the presence of the bound charge may persist because of the low temperature. These interface charges physically away from the mobile carriers in the superconducting electrodes will not be screened (as the surface bound charge will be should the tunnel barrier be a ferroelectric layer) and may affect the occupation number of the localized states, thereby redistributing the singly occupied localized states facilitating the negative Josephson couplings and thus the JD effect.

\section{Conclusion and discussion}

In conclusion, we fabricated and measured JTJs of TiN/Al$_2$O$_3$/Hf$_{0.8}$Zr$_{0.2}$O$_2$/Nb prepared by CMOS-compatible techniques. The high quality of the JTJs is demonstrated by the observations of the gap features in quasiparticle tunneling and the magnitude of the Josephson coupling close to the A-B limit as well as their responses to the magnetic fields. The JD effect was observed in these JTJs in zero magnetic fields with a large diode efficiency not seen previously in other JTJs reported thus far. Importantly, the polarity of the JD effect is controllable simply by an electric training current.

The JD effect is attributed to the coexistence of positive and negative local Josephson couplings, with the latter resulting from indirect tunneling, which leads to the spontaneous breaking of the TRS needed for the presence of JD effect. Negative Josephson coupling was also invoked in order to explain the JD effect seen in twisted high-T$_\mathrm{c}$ 2D crystals \cite{zhao_time-reversal_2023,zhu_presence_2021}, attributing the negative Josephson coupling to the $d$-wave pairing in the high-T$_c$ cuprates. The JD effect in those systems was tunable, but in a scheme more complex than that employed in the current work. 
Importantly, the physics of the trainable nonreciprocal behavior seen in the present work is fully captured by the RCSJ model with high accuracy rather than qualitatively as done in the previous work.

The demonstration of the JD prepared by CMOS-compatible ALD processes should inspire further work on technological fronts, achieving perhaps the long-sought ferroelectric JTJs as well. Even with the JTJ featuring a composite but nonferroelectric tunnel barrier used in the current work, fundamental questions are raised on this model system described by Eq.\,5, in particular, whether other novel physical phenomena in addition to the JD effect will emerge. 

\section{Acknowledgments}
This work is supported by the US Department of Energy, Office of Science, Office of Basic Energy Sciences Energy Frontier Research Centers program under Award Number DE-SC0021118. Y.L. acknowledge useful discussion with Profs. Kai Ni, Tony Leggett, and Allen Goldman.


\bibliography{HZO1}

\end{document}